\documentclass[aps,prl,reprint,groupedaddress,showpacs]{revtex4-1}
\pdfoutput=1

\usepackage{graphicx}
\usepackage{dcolumn}% Align table columns on decimal point
\usepackage{bm}% bold math
\usepackage{color} % Include color
\usepackage{amssymb} % For various other symbols

\newcommand{\lde}{\lambda_{\mathrm{De}}}
\newcommand{\lvac}{\lambda_0}

\newcommand{\be}{\begin{equation}}
\newcommand{\ee}{\end{equation}}
\newcommand{\bearr}{\begin{eqnarray}}
\newcommand{\eearr}{\end{eqnarray}}

\newcommand{\Ith}{I_{\mathrm{th}}}

\newcommand{\T}[1]{\mathbf{$1$}}

\newcommand{\micronbase}{$\mu$m}
\newcommand{\micron}{\micronbase}
\newcommand{\microns}{\micronbase\ }

\newcommand{\Lint}{L_{\mathrm{INT}}}
\newcommand{\Ln}{L_n}
\newcommand{\Lhs}{L_{HS}}
\newcommand{\Lspike}{L_{\mathrm{spike}}}
\newcommand{\RSRS}{R_{SRS}}
\newcommand{\ncr}{n_{\mathrm{cr}}}
\newcommand{\imagedirectory}{.}
\newcommand{\nscram}{n_{\mathrm{scram}}}
\newcommand{\tauspike}{\tau_{\mathrm{spike}}}

\begin{document}

\preprint{LA-UR-13-22184}

\title{Control of Stimulated Raman Scattering in the Strongly Nonlinear and Kinetic Regime Using Spike Trains of Uneven Duration and Delay: STUD Pulses}

\author{B. J. Albright,$^1$ L. Yin,$^1$ and  B. Afeyan$^2$}
\affiliation{$^1$Los Alamos National Laboratory, Los Alamos, New Mexico 87545, USA\\ 
$^2$Polymath Research Inc., Pleasanton, California 94566, USA}
\date{\today}

\begin{abstract}
Stimulated Raman scattering (SRS) in its strongly nonlinear, kinetic regime is controlled by a technique of deterministic, strong temporal modulation and spatial scrambling of laser speckle patterns, called Spike Trains of Uneven Duration and Delay (STUD pulses) [B. Afeyan and S. H\"uller, Phys. Rev. Lett. (submitted)]. Kinetic simulations show that use of STUD pulses may decrease SRS reflectivity by more than an order of magnitude over random-phase-plate (RPP) or induced-spatial-incoherence (ISI) beams of the same average intensity and comparable bandwidth.
\end{abstract}


\maketitle

Laser-plasma instabilities (LPI)   pose a  risk to the realization of laser-driven inertial confinement fusion (ICF) ignition~\cite{SanRamon}. 
The present 
approach is to use continuous, ns-time-scale illumination of a target with 
high-intensity laser beams. However, because of LPI, this  may prove to be less than ideal 
when compared with a novel technique invented by Afeyan~\cite{Afeyan13}
employing intermittent, scintillating, space-time illumination which may 
significantly reduce the levels of nonlinear optical processes. 
The efficacy of this technique, which employs Spike Trains of Uneven Duration and Delay (STUD pulses),
 has been demonstrated in  the fluid regime of instability evolution 
from low to moderate gains per speckle, where the linear growth 
is halted by the use of STUD pulses 
and any saturation is from 
pump depletion~\cite{Afeyan12,Huller12,Afeyan13}. 
This Letter focuses  on 
the application of STUD pulses to Stimulated Raman Scattering (SRS) in settings where kinetic 
nonlinearity dominates the evolution of driven electron plasma waves (EPW) 
 and where multi-laser-speckle, cooperative behavior can proceed 
 through the exchange of  
 hot electrons 
 and SRS scattered light among laser speckles~\cite{Yin_2012_PRL,Yin_PoP_2012,Yin_POP_2013}. 
 We find from our initial study that more than order-of-magnitude reduction in SRS reflectivity can be achieved. 
The key is to keep SRS growth below the level where secondary, nonlinear processes causing
 cooperative behavior among hot spots 
can occur, thus disallowing the  self-organized state.  
 
SRS is the resonant, three-wave coupling of a light wave 
into scattered light  and electron plasma waves. 
It is an LPI process occurring in large amounts in indirect-drive ICF experiments with potentially
 deleterious effects,  including scattering of laser energy out of the hohlraum, redirection of  energy within the hohlraum, and  generation of hot electrons that 
may contribute to capsule preheat. 
At the National Ignition Facility (NIF),   
 experiments  show  $\sim$50$\%$ inner-cone beam energy loss to SRS~\cite{SanRamon}. 
  Laser facilities such as OMEGA and the NIF 
 employ beam smoothing, whereby random phase plates (RPP) break  the laser beams into  ``speckles''  
 to effect a quasi-uniform (on the large scale)  intensity profile across the beam, though
 introducing into the beam small-scale,  high-intensity variations or ``speckles.'' 
  In vacuum,   speckles have  characteristic size $8 f^2 \lvac$ (longitudinally) by $ f \lvac$ (transversely), where $f$ is 
 the optical focal parameter and $\lvac$ is the laser
wavelength. The scaling of SRS reflectivity $\RSRS$ with 
 laser intensity $I$ in a solitary  speckle in plasma has been measured~\cite{Montgomery_2002_POP}
 and  found in the electron trapping regime $k \lde \gtrsim 0.3$
($k$ is the EPW wave number and $\lde = \sqrt{k_B T_e / 4 \pi n_e e^2}$ is the Debye length 
for plasma of electron density $n_e$ and temperature $T_e$) 
to behave nonlinearly, increasing sharply  at a threshold  intensity $\Ith$ and saturating  for $I > \Ith$. 
The  physics in this regime is governed by the growth of  
large-amplitude EPW that trap ``resonant'' electrons with speeds along the wave propagation direction
matching the wave's phase speed; this reduces local Landau damping~\cite{ONeil_1965}, 
enhances instability growth, and 
lowers the EPW frequency~\cite{Morales_ONeil_1972}.
At high intensity, trapping introduces variation in EPW phase velocity 
across the speckle and 
causes  wave phase fronts to bend~\cite{Yin_2007_PRL,Yin_2009,Rousseaux_2009,Banks_2011}. As EPW grow to large amplitude, 
 secondary, nonlinear processes have been proposed to 
 break the phase fronts 
into small-transverse-scale filaments~\cite{Rose_2005,Rousseaux_2009,MassonLaborde_2010_PoP}  
that further contribute to nonlinear saturation. 
An effect of   saturation, observed in simulations~\cite{Yin_2012_PRL}, 
is the generation of hot electrons and back- and side-scattered light waves that propagate obliquely out of hot spots 
and enhance SRS growth in neighboring speckles through larger SRS seed levels and reduced EPW Landau damping.   
At  high gain in two spatial dimensions (2D), this coupling enables
 networks of speckles to 
self-organize~\cite{Yin_2012_PRL} and exhibit emergent behavior where 
reflectivity exceeds that of the
sum of contributions from 
 individual, non-interacting speckles. 
The nonlinear nature of SRS in this regime is robust, with a threshold  at  modest laser 
intensity, $\gtrsim 10^{14}$~W/cm$^2$ for NIF laser conditions 
where $k \lde  \approx 0.3$  and the highest levels of backscatter are found~\cite{Kirkwood_2011}.
 
Our intent here
is to show that the use of STUD pulses~\cite{Afeyan13},  effective for 
controlling LPI over long time scales 
in the fluid regime of instability evolution~\cite{Afeyan12,Huller12,Afeyan13}, 
 may also inhibit EPW growth in the highly nonlinear, kinetic regime. 
STUD pulses deliver laser power in a sequence of  pulses or spikes on the 
instability growth or hot spot crossing time scale (sub-ps, typically, for SRS) with randomized 
laser speckle patterns in 
between one or more successive spikes. 
By introducing on-off sequences of pulses and by spatially scrambling  locations of  hot spots,  
 reinforcing processes within a hot spot and the interconnectivity between hot spots that leads to large 
instability growth are disrupted. 
STUD pulses  introduce degrees of freedom that can be optimized~\cite{Afeyan13}. These include the ratios 
$\Lhs:\Lint:\Lspike$, where the interaction length is 
$\Lint= 4 L_n [ \alpha^2 I_{14} + \left(\nu_{2} / \omega_{2}\right)^2]^{1/2}.$ Here, the density scale length is $\Ln=| \nabla \log n_e |^{-1}$, 
$I_{14}$ is intensity in $10^{14}$~W/cm$^2$, 
$\nu_{2}$ and $\omega_{2}$ are  the local Landau damping rate and frequency of the EPW, respectively, 
 $\alpha^2 = 1.42 \lambda_{0}^{[\mu m]} \Ln^{[100\ \mu m]} \left| V_1/V_2\right| (n_e/\ncr)^{-1}$, and  
$V_1/V_2$ is the ratio of group velocities of scattered SRS light to the EPW~\cite{Afeyan13}. 
 Spike length $\Lspike$ is the
distance traveled by scattered light during the `on' time $\tauspike$ 
and $\Lhs \sim 4 f^2 \lambda_0 = 90$~\microns is the characteristic size of a hot spot in our plasma. Other degrees of freedom are  
duty cycle (the ratio of $\tauspike$ to on+off time),
 the spatial scrambling rate $\times \nscram$ (how many identical spikes before the RPP pattern changes) and the ``jitter'' (random variation 
 of each $\tauspike$; all calculations  in this Letter used $0\%$ jitter). 
Hence,  ``5000$\times$1, 1:1:1/2'' indicates a STUD pulse sequence with 50\% duty cycle, $0\%$ jitter, 
and a spike duration half as long as a hot spot crossing time in plasma where the 
time to cross $\Lint$ for the three-wave process is comparable to that of crossing the hot spot. 
Most of the results  we present are for cases 
 5000$\times$1,
 1:1:1/2 and  1:1:1  
 in strong to very 
strong nonlinear kinetic regimes (SRS gains of 6--11 at the average intensity). 
Note that configurations where `on' time is much greater than `off' time, e.g., 8000$\times$1 or 9500$\times$1, 
 resemble the ISI model of beam smoothing~\cite{Lehmberg00} at the same bandwidth. 
 
%----------------------------------------------------------------------
% Fig. 1
\begin{figure}
\includegraphics[width=90mm]{\imagedirectory/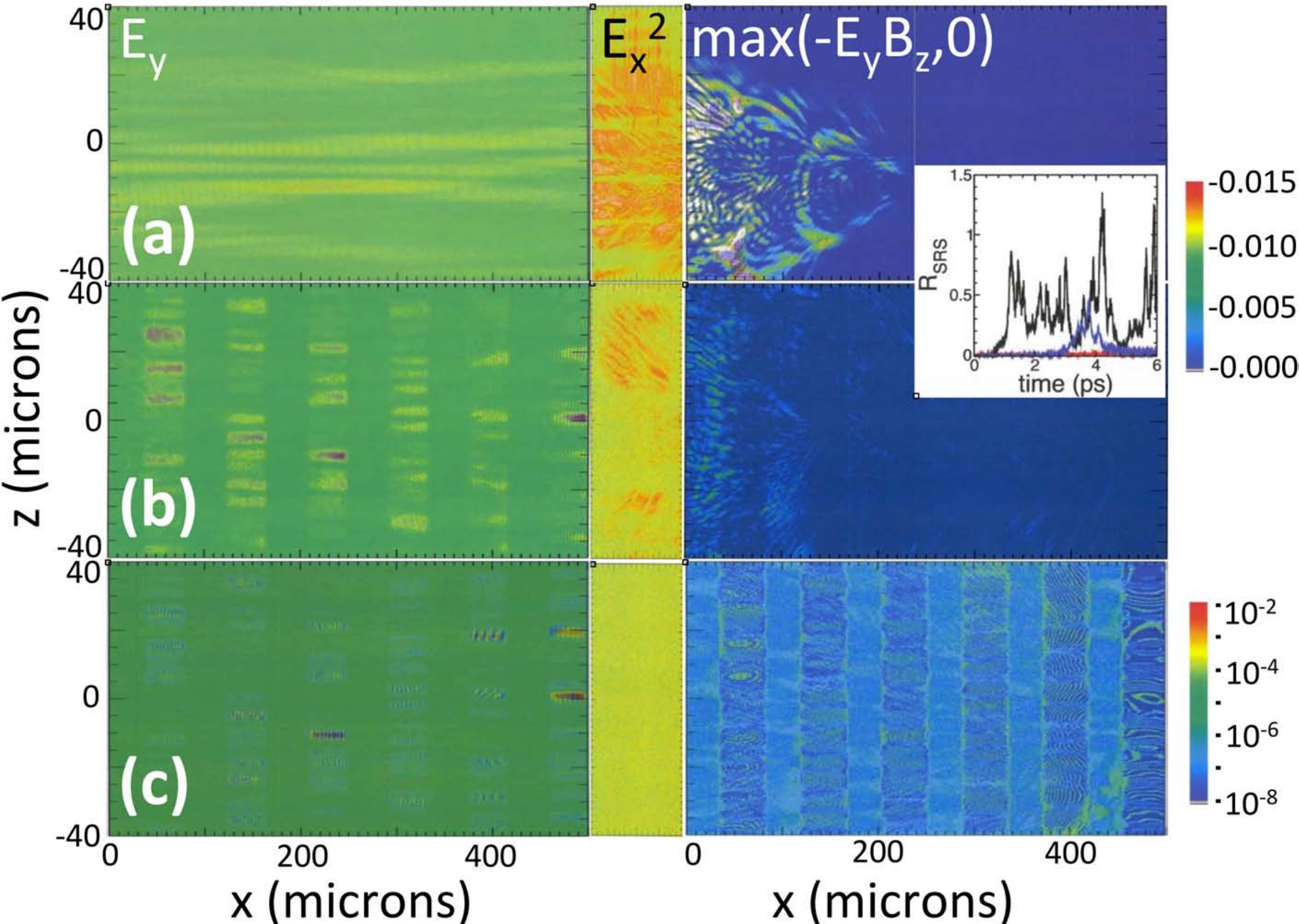}
\caption{(Color)
\footnotesize
$E_y$ (left) and corresponding instantaneous back-scattered Poynting flux $\mathrm{max}\left(-E_y B_z, 0\right)$ (right) over the 2D simulation volume 
for three cases: (a) an RPP laser beam at average laser intensity $\left< I\right> = 5\times 10^{14}$~W/cm$^2$ (top), (b) a 5000$\times$1, 1:1:1/2 STUD pulse 
beam at time-averaged incident laser intensity $\left< I\right> = 5\times 10^{14}$~W/cm$^2$ (center); and, (c) the same STUD pulse beam, but 
with $\left< I\right> = 3.2\times 10^{14}$~W/cm$^2$ (bottom) (note logarithmic scale on Poynting flux). The center panels are electrostatic field energy $E_x^2$ for the leftmost 80~\microns of
each simulation, showing EPW wave amplitude correlated with instantaneous SRS backscattered Poynting flux.  
The inset is reflectivity vs. time for cases (a) (black) and (b) (blue); (c) (red) evinces negligible backscatter. 
The times shown are 1.6 ps (a) and 3.6~ps (b,c), chosen 
when  large, bursts of SRS backscatter were present in  (a) and (b). 
}
\label{fig:fig1}
\end{figure}
%----------------------------------------------------------------------

To understand better the behavior of STUD pulses in the trapping regime,  
we have run VPIC particle-in-cell simulations~\cite{Bowers_2008}
of a 2D plasma of  size $500\times 80$~\microns in $(x,z)$, with 
a  laser beam polarized along $y$ launched at $x=0$ as described in Ref.~\cite{Yin_PoP_2012}.
The laser has wavelength $\lambda_0 = 0.351$~\microns and an RPP speckle 
 pattern for $f\backslash 8$ speckles is used, approximating a NIF inner-cone 
 beam in  hohlraum  plasma. 
 The density has a gradient  along $x$ with  
 $n_e = 0.12 \ncr$ at the center, changing from $\pm 0.013 \ncr$ to 
$\pm 0.03 \ncr$ across the box, comparable to the $\Ln \sim \mathrm{mm}$ 
encountered in NIF ignition hohlraums 
 in regions of high SRS backscatter~\cite{Kirkwood_2011}. 
Taking $\nu_2 = \nu_2^{\mathrm{Max}}$, as for Maxwellian plasma, 
 in the $k \lde \approx 0.3$ regime yields
$\Lint \sim 95$--$99$~\microns for the range of intensities simulated.
 We use $36864\times 4096$ cells ($ \Delta x = 1.2 \lde$ and $\Delta z = 1.7 \lde$) and $256$--$512$ 
 electron macroparticles/cell; ions are stationary. 
 %(Statistical noise levels are comparable to those in 
%physical NIF hohlraum plasma~\cite{Yin_PoP_2012}.) 
The electrons have $T_e = 2.6$~keV ($k\lde = 0.3$). 
The STUD pulse speckle patterns are generated from pre-computed 
RPP phases for a % $\sim$1~cm, 
wide beam,  
sampling 80-\micron,  non-overlapping segments for each STUD pulse.
 Each simulation employs the same
sequence of speckle patterns to within an overall intensity  modulation, allowing variation of intensity, duty cycle, and 
modulation period. (Statistical variation was assessed by altering the sequences of STUD pulses; 
 $\sim$10$\%$ relative $\RSRS$  variation was found in a range of cases considered.) The simulation boundaries absorb electromagnetic waves and 
 reinject electrons as Maxwellian at initial temperature $T_e$. 
  The simulations were run until apparent ``steady-state'' in time-averaged $\RSRS$, $10$--$20$~ps.

Fig.~1 shows a comparison of three simulations: (a) (top row) is for an RPP beam with $\left< I\right> = 5\times 10^{14}$~W/cm$^2$ ($G =11$); 
 (c) (bottom) is for a STUD pulse beam of  time-averaged laser intensity $\left< I \right> = 3.2\times 10^{14}$~W/cm$^2$ ($G =7.5$). Linear SRS 
 gains $G$ are computed from 
 $   G = 4 \pi \left(\gamma_0/\omega_0\right)^2 \left(2 \pi L_n / \lambda_0\right) {g(n/n_c)}^{-1} (1-\nu_1\nu_2 / \gamma_0^2)$, 
 where $\left(\gamma_0/\omega_0\right) = 0.0043 %4.267 \times 10^{-3} 
 \sqrt{I_{14}} \, \lambda_0^{[\mu m]}$, $\nu_1$ is the damping rate of the daughter light wave, 
 and  $g(n)\equiv \sqrt{1 - 2 \sqrt{n}}\left[(1/\sqrt{n})-1 \right]$~\cite{Afeyan13}.
Accounting for  backscatter loss, (a) and (c) have comparable net time-averaged  power incident on the left boundary, though (c) has only 
64\% of the incident time-averaged laser power. 
Case (b) (center) is for a STUD pulse 
beam at the same time-averaged incident laser intensity as (a): $\left< I\right> = 5\times 10^{14}$~W/cm$^2$ ($G=11$). The leftmost panels show 
$E_y$ (or the vacuum speckle pattern for the RPP case).  
The rightmost panels are instantaneous backscattered Poynting 
flux $\mathrm{max}\left(-E_y B_z, 0\right)$. Case (a) evinces continual bursts of self-organized backscatter with peak $\RSRS>1$. In  (c),  no 
self-organization is seen
in backscattered light or longitudinal electric field.  
 Case (b) is intermediate, showing quiescent periods of low backscatter punctuated by 
 occasional episodes of partial self-organization 
  when large-amplitude speckles ($I \gtrsim 10 \left<I\right>$) exhibit large-amplitude EPW and 
 secondary
 processes, such as obliquely side-scattered light, occur at sufficient amplitude to seed SRS in otherwise stable regions of plasma
(seen in the finite backscattered SRS Poynting flux across the left of the box). 
  The instantaneous $\RSRS$ at the left  boundary is 
shown in the inset for cases (a--c) (black, blue, and red curves, respectively); the times plotted are 1.6~ps for the RPP (during the first large SRS burst), and 3.6~ps 
for the STUD pulse simulations [during the first, large SRS burst  in (b)]. 
The central panels are $E_x^2$ over the leftmost 80~\microns of the  volume and indicate EPW amplitude correlated with 
the  large bursts of SRS in (a) and (b).  

%----------------------------------------------------------------------
% Fig. 2
\begin{figure}

\includegraphics[width=45mm]{\imagedirectory/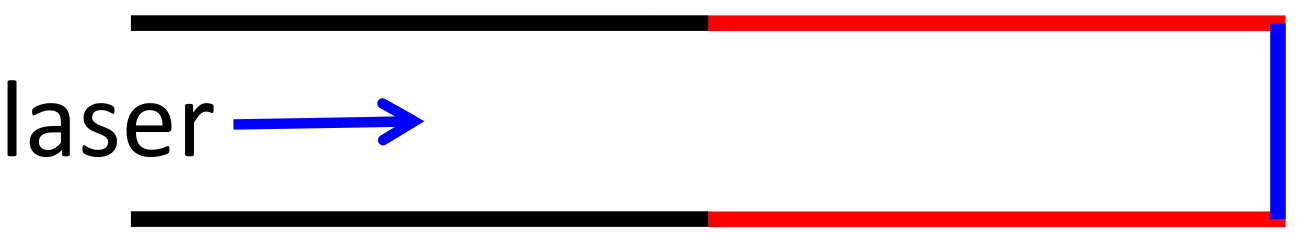}

\includegraphics[width=60mm]{\imagedirectory/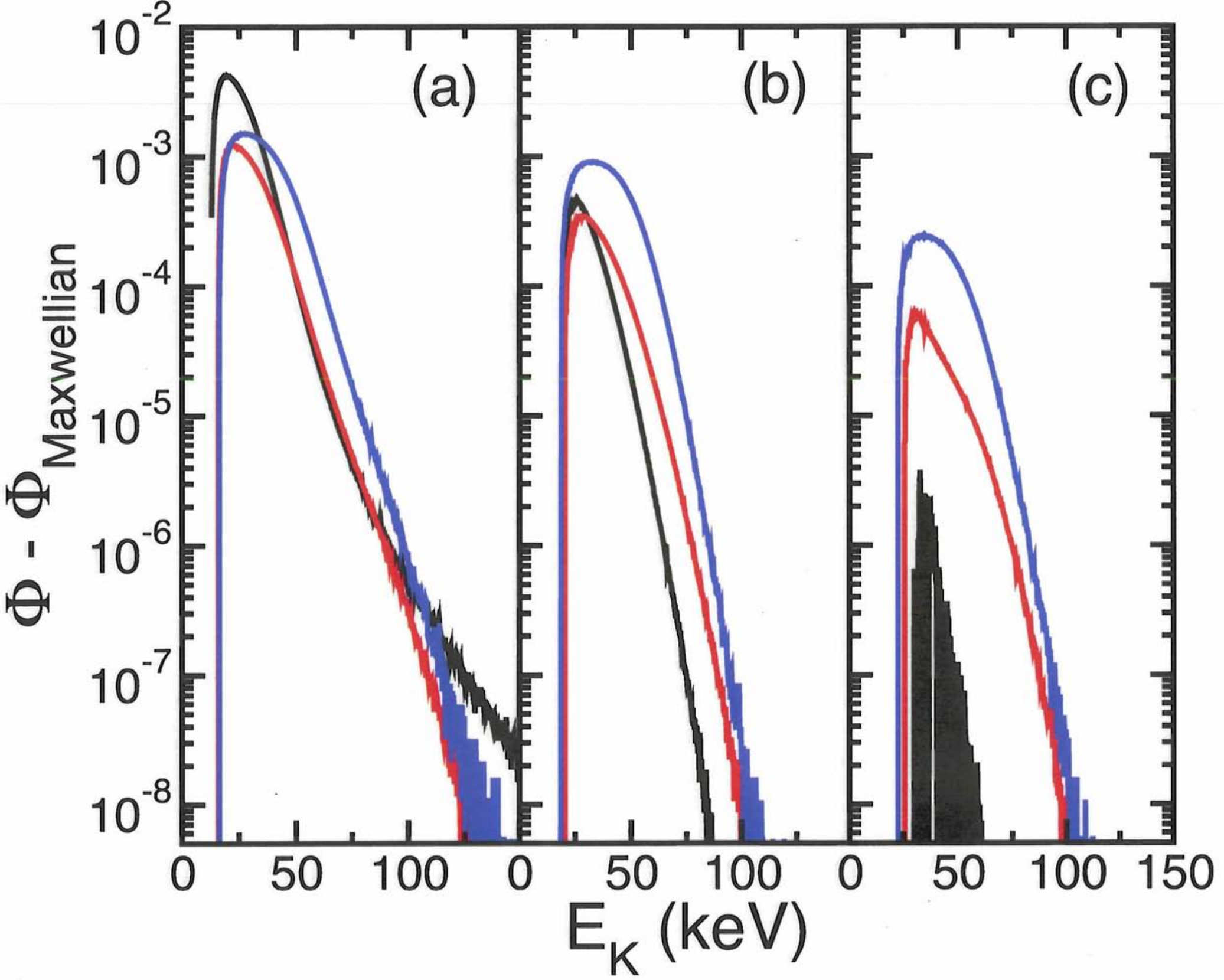}

\caption{(Color)
\footnotesize
Hot electron flux per unit length $\Phi$ vs. electron energy ($E_K$) for the three simulations 
 in Fig.~1. Shown are  trapped particle fluxes,  obtained by subtracting contributions from a Maxwellian. 
Fluxes are measured on  boundary regions, as indicated by the colors 
(c.f. the simulation box above, drawn to scale):  $z = \pm 40$~\micron, $0 < x<250$~\microns (black); $z = \pm 40$~\micron, $250 < x<500$~\microns (red) and $x = 500$~\microns (blue).
}
\label{fig:fig2}
\end{figure}
%----------------------------------------------------------------------

%----------------------------------------------------------------------
% Fig. 3
\begin{figure}
\includegraphics[width=65mm]{\imagedirectory/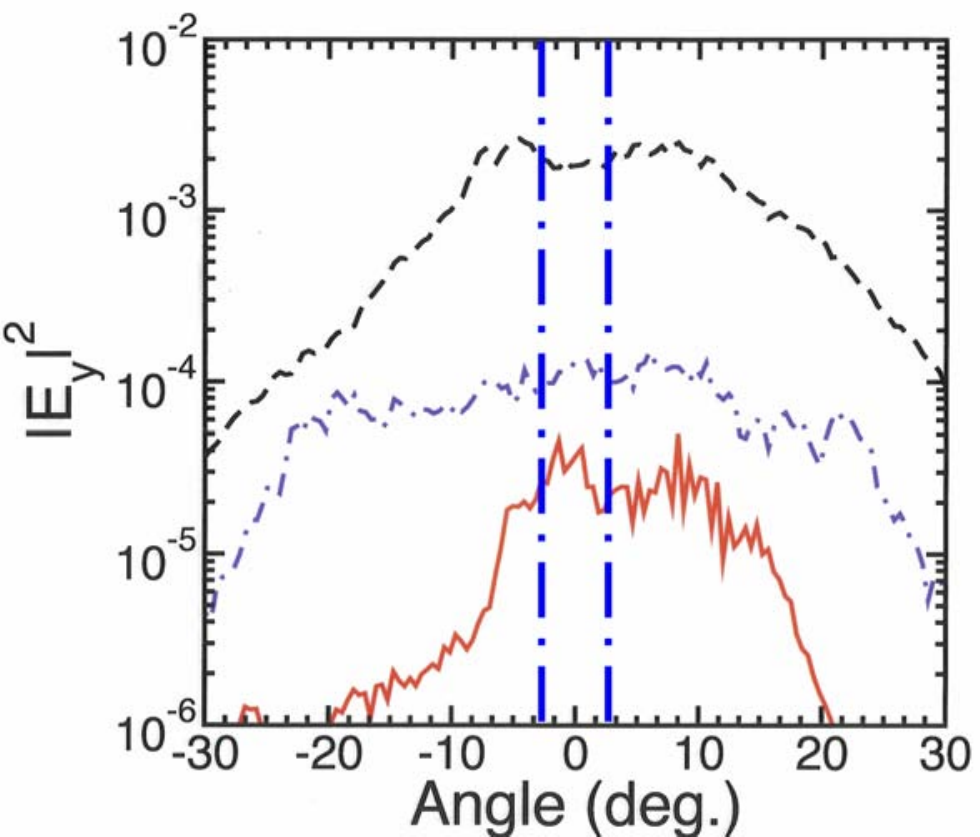}
\caption{(Color)
\footnotesize
Angular distribution of the time-averaged backscattered light power for 
case (a) (black, dashed), (b) (dot-dashed, blue), and (c) (red) as in Figs.~1 and 2. 
The spectra for (b) and (c) evince lower backscattered light power, but 
at finite angle with respect to the incident laser (cone angle $\left| \theta\right| <  1/ 2 f$, shown by the vertical lines). 
}
\label{fig:fig3}
\end{figure}
%----------------------------------------------------------------------

In Fig.~2, we compare for  cases (a--c) time-integrated hot electron flux per unit area exiting the simulation.
The black curves are fluxes leaving the $\pm z$ boundaries from the left half of the simulation volume, the red curves, 
 leaving $\pm z$ from the right  of the volume, and the blue curves, leaving from the $+z$  boundary.
Prior work showed that large fluxes of tail electrons leaving the left  of the  volume (i.e., large black curves) 
are signatures of large-amplitude EPW with ensuing nonlinear self-focusing and filamentation and, ultimately,
 collective behavior among speckles~\cite{Yin_PoP_2012,Yin_POP_2013}. 
The three cases  evince elevated distribution function tails as a consequence of trapping, though the RPP 
case traps not only far more tail electrons 
 [60$\times$ more than (c), 6$\times$ more than (b)],
 but also shows far more side-scattered hot electrons exit nearest the laser entrance; moreover, 
 hot electrons at very high energy ($E_K > 100$~keV) are present (absent for the STUD pulse beams). 
 The use of STUD pulses has therefore 
  decreased the number of hot electrons  exchanged laterally among laser speckles, a key part of 
  inter-speckle self-organization~\cite{Yin_POP_2013} and a possible contributor to capsule preheat in ICF experiments. 
 In Fig.~3, we compare the angular spread of SRS light  for the same cases as above. 
The use of STUD pulses leads to a dramatic
 overall reduction in  SRS power (and hence, amplitude of the SRS seed in neighboring speckles). 
 As with the RPP, the angular spread  is  finite, with most of the power  
falling outside the incident laser cone $\left| \theta\right| <  1/ 2 f$ shown by the vertical lines. 
While the existence of coherent, oblique cones of backscattered light is not unique to this nonlinear regime---indeed, they appear 
in paraxial models with diffraction~\cite{Kolber_1993,Afeyan13}---additional side-scatter results from
trapping and EPW filamentation~\cite{MassonLaborde_2010_PoP,Yin_2012_PRL}
 that is absent in fluid models; the use of STUD pulses reduces dramatically 
the levels of such side-scatter.

 Finally, in Fig.~4, we show the dependence of $\RSRS$ on time-averaged incident laser intensity (left) 
and linear gain at the average intensity (right) for 
  RPP and STUD pulses over a  range of plasma and laser conditions. 
The use of STUD pulses reduces dramatically $\RSRS$ compared with 
RPP and ISI-like beams with the same time-averaged laser power. 
This is so even in cases of very high linear gain. 
As seen from comparison of the $\RSRS$ from the ISI-like points (the 8000$\times$1, 1:1:1/2 and 9500$\times$1, 1:1:1/2) 
and   5000$\times$1, 1:1:1/2 cases, ``healing time'' 
is key: it is not enough to simply add bandwidth and spatial scrambling. 
By optimizing this healing time for given ``on+off'' time and time-averaged power, 
STUD pulses may  be optimized to significantly outperform ISI. 
From comparison of the 5000$\times$4, 1:1:1/2 and the 5000$\times$1, 1:1:1/2 cases, we find that  
 spatial scrambling of the locations of the hot spots is also necessary
 to avoid effects of recurrence and correlation among successive hot spots~\cite{Note2D3D}.  
Also, for fixed ``on+off'' time and time-averaged power, lengthening 'off' time 
requires shortening $\tauspike$ and increasing the average speckle intensity correspondingly.
Taken to an extreme, this can lead to enhanced trapping and associated EPW nonlinearity, 
 evidenced by the 2000$\times$1 datum shown in Fig.~4 [which also has significant hot electron sidescatter (not shown) compared with the
 5000$\times$1, 1:1:1/2 case at the same average power].

%While do not have space in this Letter to explore in detail how STUD pulses affect the nonlinear physics governing SRS coupling,  we  
%note some important 
%time and length scales: First, in the strongly nonlinear, trapping regime, EPW growth to large amplitude occurs over sub-ps intervals, so 
%STUD pulses must be modulated
%at least this rapidly to affect the dynamics. 
%Second, 
%While one may be tempted to conclude that use of STUD pulses lowers $\RSRS$  simply because the time between successive 
 %pulses allows the EPW to damp, the picture is more complex, as e
%Properly optimized STUD pulses prevent  EPW from reaching a nonlinearly saturated state before the end of the pulse and the hot 
%spot is (statistically) relocated.
Examination of velocity distribution 
functions and EPW amplitudes shows strong trapping and only modest EPW damping between
cycles. This trapping modifies $\Lint$ and suggests that 
%Third, for the most
%effective STUD pulses modeled here, 
%$t_{\mathrm{on}} \approx f \lambda_0 / \vthe$, the lateral hot electron crossing time across a laser speckle prior to the onset of  
%nonlinear self-focusing or filamentation.
 %which  comparing the inferred, 
 %comparing trapping-modified $\Lint$ with $\Lspike$ suggests 
 possible threshold behavior  may arise when $\Lspike$ becomes less than $\Lint$ and
 when SRS goes from strong to weak damping. 
 Consider 
the two 5000$\times$1 and the 9500$\times$1 STUD pulse cases at the highest
 intensity ($G=11$).  In the former, SRS in the largest amplitude hot spots ($I \gtrsim 10 \left<I\right>$) 
 would  be in the 
 weak damping limit (WDL) if one applies the inferred $\nu_{2}$ from simulations ($\approx 0.1 \nu_2^{\mathrm{Max}}$), 
 and $\Lint \approx 120$~\micron.  
 The 1:1:1/2 STUD pulse case, with the lowest $\RSRS$, has $\Lspike \sim 0.37 \Lint$ for these maximal 
 speckles, i.e.,   STUD pulses  much shorter than $\Lint$. In contrast, the 1:1:1 case has 
 $\Lspike \sim  \Lint$. In the 9500$\times$1, 1:1:1/2 ISI-like case, while $\Lint = 0.5 \Lspike$, 
  reduction of $\nu_{2}$  causes the SRS to go to the WDL for an average intensity speckle, with correspondingly large $\RSRS$. 

%----------------------------------------------------------------------
% Fig. 4
\begin{figure}
\includegraphics[width=80mm]{\imagedirectory/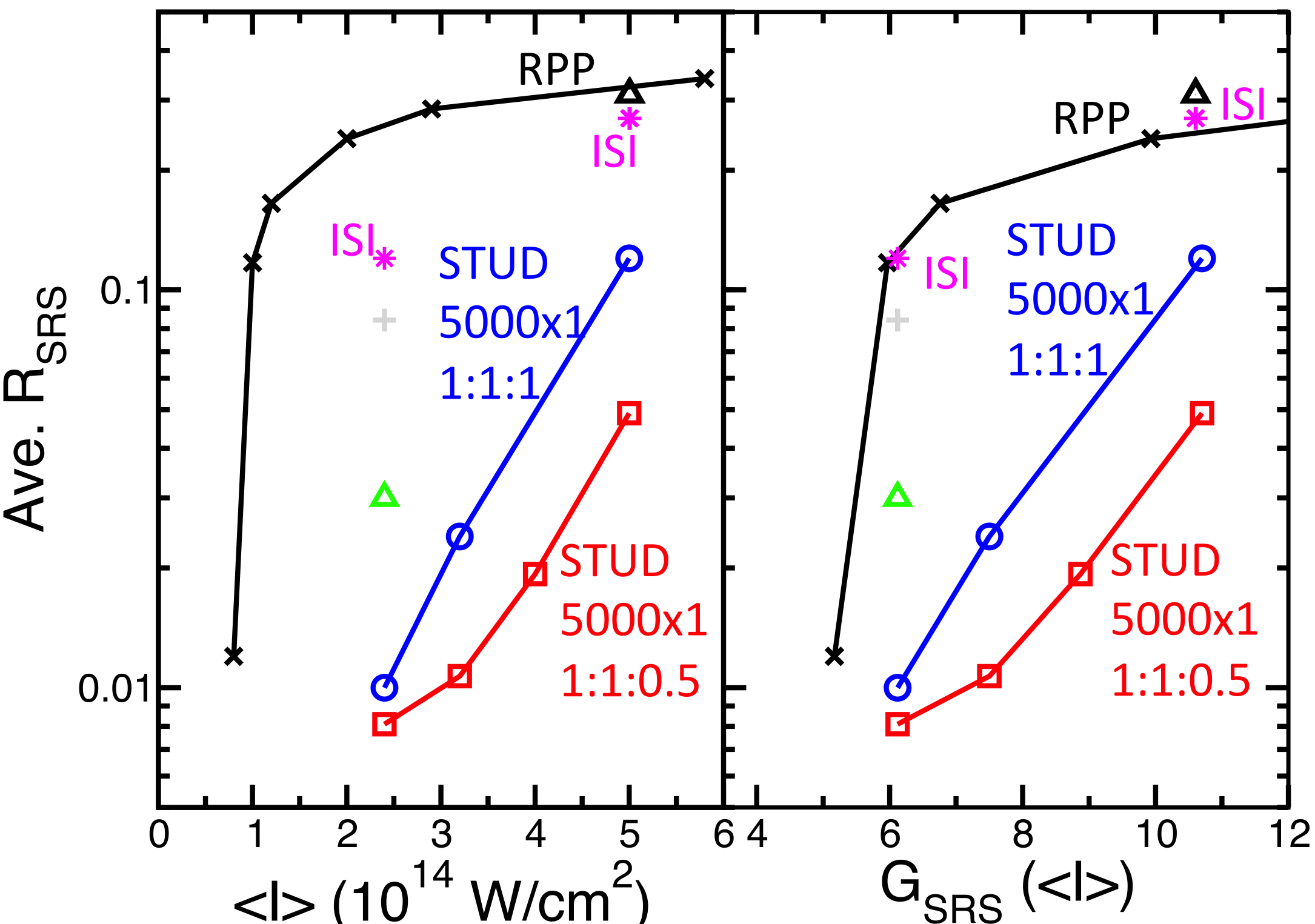}
\caption{(Color)
\footnotesize
SRS reflectivity  plotted as a function of average incident intensity (left) and linear gain (right) for RPP and STUD-pulse beams for a variety of
laser and plasma conditions, showing the efficacy of STUD pulses for reducing $\RSRS$. 
The black points (marked $\times$ and $\triangle$) are for RPP beams for two different density variations across the box 
in $x$: $\pm 0.013 \ncr$ and $\pm 0.03 \ncr$, 
respectively. The red   ($\square$) and blue ($\circ$) curves are for STUD pulses with 5000$\times$1, 1:1:1/2 and 5000$\times$1, 1:1:1 
(i.e., twice the ``on'' time) for the latter density profile. 
The magenta datapoints labeled ``ISI'' are
for STUD pulses with 
8000$\times$1, 1:1:1/2 (left) and 
9500$\times$1, 1:1:1/2 (right) and indicate little advantage over RPP for the conditions shown. 
The green ($\triangle$) is for STUD pulses with 5000$\times$4, 1:1:1/2, i.e.,  scrambling  every 4 pulses. The ($+$)  is for 
%a very 
%short duty cycle case: 
2000$\times$1, 1:1:1/2. 
}
\label{fig:fig4}
\end{figure}
%--------------------------------------------------------------------

We have shown that SRS reflectivity may be lowered by more than an order of magnitude
 with the use of properly designed STUD pulses in settings where 
 EPW trapping-induced nonlinearity is prevalent. This reduction stems from arresting
  large-amplitude EPW that give rise to
  cooperative behavior among laser speckles through the exchange of hot electrons and backscatter SRS waves, 
  thus disallowing their self-organization. 
The substantial promise and generality of the STUD pulse technique~\cite{Afeyan13}
to a range of settings, including SRS in the strongly nonlinear, trapping regime considered here, 
would seem to impel serious
consideration for how STUD pulses might be achieved 
in future ICF laser systems such as the Green option on the NIF 
or next-generation high-repetition-rate laser systems.  
  
Work performed under the auspices of the U.S. Dept. of Energy under contract W-7405-ENG-36. 
BJA \& LY  were 
supported by DOE NNSA ICF and LDRD Programs and the DOE NNSA-OFES Joint program in HEDLP. 
BA's work was supported by grants 
from the DOE NNSA-OFES Joint program in HEDLP and by SBIR grants from OFES. 
 Simulations were performed on ASC Roadrunner and Cielo. 
We acknowledge 
useful discussions with S. H\"uller, J. Garnier, J. Fern\'andez, D. Montgomery, J. Kline and S. Batha.


\begin{thebibliography}{99}


\bibitem{SanRamon}
William Goldstein and Robert Rosner, eds., Proc. Science of Fusion Ignition on NIF Workshop, LLNL-TR-570412,
San Ramon, CA, Lawrence Livermore National Security, LLC, (2012);
William L. Kruer in \textit{Laser-Plasma Interactions and Applications}, edited by P. McKenna et al., Scottish Graduate Series (Springer, Switzerland, 2013). 

\bibitem{Afeyan13}
B. Afeyan and S. H\"uller, ``Optimal Control of Laser-Plasma Instabilities using Spike Trains of Uneven Duration and Delay (STUD Pulses)," 
Phys. Rev. Lett. (submitted); also arXiv:1304.3960. 

\bibitem{Afeyan12} 
B. Afeyan and S. H\"uller, Europ. Phys. J. Web of Conferences (in press); also arXiv:1210.4462.

\bibitem{Huller12}
S. H\"uller and B. Afeyan, Europ. Phys. J. Web of Conferences (in press); also arXiv:1210.4480.

\bibitem{Yin_2012_PRL} 
L. Yin, B. J. Albright, H. A. Rose, K. J. Bowers, B. Bergen and R. K. Kirkwood, Phys. Rev. Lett. \textbf{108}, 245004 (2012).

\bibitem{Yin_PoP_2012}
L. Yin
B. J. Albright,
H. A. Rose,
K. J. Bowers,
B. Bergen,
R. K. Kirkwood,
D. E. Hinkel,
A. B. Langdon,
P. Michel,
D. S. Montgomery,
J. L. Kline,
Phys. Plasmas {\bf 19}, 056304 (2012).

\bibitem{Yin_POP_2013}
L. Yin, B. J. Albright, H. A. Rose, D. S. Montgomery, J. L. Kline,
R. K. Kirkwood, P. Michel, K. J. Bowers, B. Bergen,
Phys.\ Plasmas {\bf 20}, 012702 (2013).

\bibitem{Montgomery_2002_POP}
D. S. Montgomery, J. A. Cobble, J. C. Fern{\'a}ndez, R. J. Focia, R. P. Johnson, R. Renard-LeGalloudec, 
H. A. Rose, and D. A. Russell, Phys. Plasmas \textbf{9}, 2311 (2002). 

\bibitem{ONeil_1965}
T. O'Neil,
Phys. Fluids {\bf 8}, 2255 (1965).

\bibitem{Morales_ONeil_1972}
G. J. Morales,
and T. M. O'Neil,
Phys. Rev. Lett. {\bf 28}, 417 (1972).

\bibitem{Yin_2007_PRL}
L. Yin, B. J. Albright, K. J. Bowers, W. Daughton, and H. A. Rose, Phys. Rev. Lett., 99, 265004 (2007).

\bibitem{Yin_2009}
L. Yin,
B. J. Albright,
H. A. Rose,
K. J. Bowers,
B. Bergen,
D. S. Montgomery,
J. L. Kline,
and J. C. Fern\'{a}ndez,
Phys. Plasmas {\bf 16}, 113101 (2009).

\bibitem{Banks_2011}
J. W. Banks, R. L. Berger, S. Brunner, B. I. Cohen, and J. A. F. Hittinger, 
Phys. Plasmas \textbf{18}, 052102 (2011).

\bibitem{Rousseaux_2009}
C. Rousseaux, S. D. Baton, D. B\'enisti, L. Gremillet, J. C. Adam, A. H\'eron, D. J. Strozzi, and F. Amiranoff, 
Phys. Rev. Lett. \textbf{102}, 185003 (2009). 

\bibitem{Rose_2005}
H. A. Rose,
Phys. Plasmas {\bf 12}, 12318 (2005); H. A. Rose and L. Yin,
Phys. Plasmas {\bf 15}, 042311 (2008).

\bibitem{MassonLaborde_2010_PoP}
P. E. Masson-Laborde, W. Rozmus, Z. Peng, D. Pesme, S. H\"uller, M. Casanova, V. Yu. Bychenkov, T. Chapman, and P. Loiseau, 
Phys. Plasmas {\textbf{17}}, 092704 (2010). 

\bibitem{Kirkwood_2011}
R. K. Kirkwood et al., 
%P. Michel, R. London, J. D. Moody, E. Dewald, L. Yin, J. Kline, D. Hinkel, D. Callahan, N. Meezan, E. Williams, L. Divol, B. J. Albright, K. J. Bowers, E. Bond, H. Rose, Y. Ping, T. L. Wang, C. %Joshi, W. Seka, N. J. Fisch, D. Turnbull, S. Suckewer, J. S. Wurtele, S. Glenzer, L. Suter, C. Haynam, O. Landen, and B. J. Macgowan,
Phys. Plasmas {\bf 18}, 056311 (2011);
D. E. Hinkel et al., 
%M. D. Rosen, E. A. Williams, A. B. Langdon, C. H. Still,D. A. Callahan,J. D. Moody,P. A. Michel,R. P. J. Town,R. A. London,and S. H. Langer,
Phys. Plasmas {\bf 18}, 056312 (2011); 
S. H. Glenzer et al., Phys. Rev. Lett. \textbf{106}, 085004 (2011). 

\bibitem{Lehmberg00}
R. Lehmberg and J. Rothenberg, J.Appl. Phys. {\bf 87}, 1012 (2000).

\bibitem{Bowers_2008}
K. J. Bowers, B. J. Albright, L. Yin, B. Bergen, and T. J. T. Kwan,
Phys. Plasmas \textbf{15}, 055703 (2008);
K. J. Bowers, B. J. Albright, B. Bergen, L. Yin, K. J. Barker, and D. J. Kerbyson,
Proceedings of the ACM/IEEE conference on Supercomputing, Austin, 2008
(Piscataway, NJ, USA: IEEE Press) pp 1-11;
K. J. Bowers, B. J. Albright, L. Yin,
W. Daughton, V. Roytershteyn, B. Bergen, and T. J. T. Kwan,
J. Phys.: Conf. Ser. \textbf{180}, 012055 (2009).

\bibitem{Kolber_1993}
T. Kolber, W. Rozmus, and V. T. Tikhonchuk, Phys. Fluids B, \textbf{5}, 138 (1993). 

\bibitem{Note2D3D}
 Because of how geometry affects multi-speckle coupling, transverse cross-speckle coupling 
through SRS side-loss hot electrons has been shown to be lower in 3D vs. 2D~\cite{Yin_POP_2013}. 
Moreover, the probability of spatial recurrence of 
hot spots is vastly smaller in 3D than in 2D. 
As a consequence,  
the advantages of STUD pulses in this nonlinear kinetic regime of SRS 
are expected to be more pronounced in more realistic, 3D geometry. 
\end{thebibliography}
\end{document}